\documentclass[aps,prl,preprint,superscriptaddress]{revtex4}

\usepackage{graphicx}
\usepackage{bm}
\usepackage{amsmath}
\usepackage{amssymb}
\usepackage{hyperref}

\begin{document}

\title{Deflection of suspended graphene by a transverse electric field}

\author{Zhao Wang}
\email{wzzhao@yahoo.fr}
\affiliation{EMPA - Swiss Federal Laboratories for Materials Testing and Research, Feuerwerkerstrasse 39, CH-3602 Thun, Switzerland}

\author{Laetitia Philippe}
\affiliation{EMPA - Swiss Federal Laboratories for Materials Testing and Research, Feuerwerkerstrasse 39, CH-3602 Thun, Switzerland}

\author{Jamil Elias}
\affiliation{EMPA - Swiss Federal Laboratories for Materials Testing and Research, Feuerwerkerstrasse 39, CH-3602 Thun, Switzerland}

\begin{abstract}
We investigate the electromechanical response of doubly clamped graphene nanoribbons to a transverse gate voltage. An analytical model is developed to predict the field-induced deformation of graphene nanoribbons as a function of field intensity and graphene geometry. This model is validated thought atomistic simulations using the combination of a constitutive charge-dipole model and a pseudo-chemical many-body potential. As a newly observed effect of electric polarization, this field-induced deflection allows the graphene to oscillate at its natural frequency, which is found to decrease dramatically with increasing graphene size.
\end{abstract}

\maketitle
If we bring a glass rod electrically charged by rubbing with silk near a hair, the hair will be attracted to the rod. Here we demonstrate a similar electrostatic effect occurring in a graphene nanoribbon, a one-atom-thick carbon crystalline layer, which has been shown to have interesting electronic properties \cite{CastroNeto2009,Beenakker2008} at nanoscale. In particular, graphene's electronic gap tunable in external electromagnetic fields makes it desirable for numerous applications in nanoelectronic devices \cite{Novoselov2004,Son2006b,Yang2007c,Han2007a,Castro2007,Novikov2007a,Yan2007a}. In such devices, the graphene is usually suspended between supports and exposed to an external electric field \cite{Bunch2007}. Since electronic transport properties of graphene are highly sensitive to the change of atomic structure \cite{Son2006,Han2007a}, understanding its mechanical behaviors in response to an applied electric field is a crucial part for the development of graphene-based electronic and electromechanical devices. 

In recent experiments, vibrations induced by an electric field were observed in carbon nanotubes (CNTs) \cite{Poncharal-99}. This property has then been exploited in the design of a number of CNT-based nanoelectromechanical devices \cite{Anantram2006}. It would not be surprising if graphene exhibits similar structural instability in an electric field as nanotubes do \cite{Wang200903}, in view of the large similarity in their in-plane electric polarizabilities \cite{kozinsky-06} and mechanical properties \cite{Lee2008}. Indeed, as an extremely thin membrane with well-defined electronic properties and low lateral stiffness, a suspended graphene sheet should be an ideal material for use in an electromechanical oscillator or resonator.  

When an electric field is applied across a thin neutral molecule, a moment of force acting on the molecule will arise as a result of electric polarization. It bends the molecule to the field direction \cite{joselevich-02}. By such a deflection, electrical potential energy is converted and stock in the molecular structure. The mechanism of this energy conversion relies on the interaction between the electric field and the polarized charges on the molecule. This energy can be released in a form of mechanical oscillation by removing the field as far as the deflection of the molecule is elastic \cite{Garcia-Sanchez2008}.  

A natural question to ask at this point is, \textit{how does the shape of graphene change in response to applied transverse electric fields?} In this work, we present an analytical approach to predict deformations of graphene nanoribbons (GNRs) by an electric field, demonstrating the coupling between the electric polarization and the mechanical resistance of graphene. This approach is validated through molecular simulations in view of the difficulty to establish an experimental quantification system of nanoscale electric polarization effects. Resonance frequency of GNRs with different length is measured by simulating vibration of GNRs by means of molecular dynamics (MD).  

We performed molecular simulations \cite{Wang200903} to compute the deflection of suspended GNRs by an electric field, minimizing the total potential energy of the system, which consists of two terms: an internal potential due to the C-C chemical bonds $U_{c}$, and an external potential $U_{e}$ arising from the interaction with an external electric field. $U_{c}$ is calculated using the adaptive interatomic reactive empirical bond order (AIREBO) potential function \cite{Stuart2000a}, which has been used in recent studies on mechanical properties of CNTs \cite{Ni2002} and GNRs \cite{Shenoy2008}. $U_{e}$ is computed using a constitutive Gaussian-regularized electrostatic model, in which each atom is modeled as an induced dipole $\bm{p}$ and a quantity of free charge $q$ \cite{mayer-07-01},

\begin{multline}
U_{e}=\sum_{i=1}^N{q_i(\chi_i+V_i)}-\sum_{i=1}^N{\bm{p}_i\cdot\bm{E_i}}+
\frac{1}{2}\sum_{i=1}^N{\sum_{\substack{j=1}}^N{q_i T^{i,j}_{q-q} q_j}}\\-\sum_{i=1}^N{\sum_{\substack{j=1}}^N{\bm{p}_i \cdot \bm{T}^{i,j}_{p-q} q_j}}
-\frac{1}{2}\sum_{i=1}^N{\sum_{\substack{j=1}}^N{\bm{p}_i \cdot \bm{T}^{i,j}_{p-p} \cdot \bm{p}_j}}
\end{multline}

\noindent where $N$ is the total number of atoms, $\chi$ is the electron affinity, $V$ and $\bm{E}$ stand for the external potential and electric field, respectively. $T$ and $\bm{T}$ are the electrostatic interacting tensors. This model has recently been validated through electrostatic force microscopy experiments on CNTs \cite{zhaowang-08-01}. Further details about the simulation can be found in Refs.\cite{zhaowang-07-03,Wang200903}. 

\begin{figure}[ht]
\centerline{\includegraphics[width=13cm]{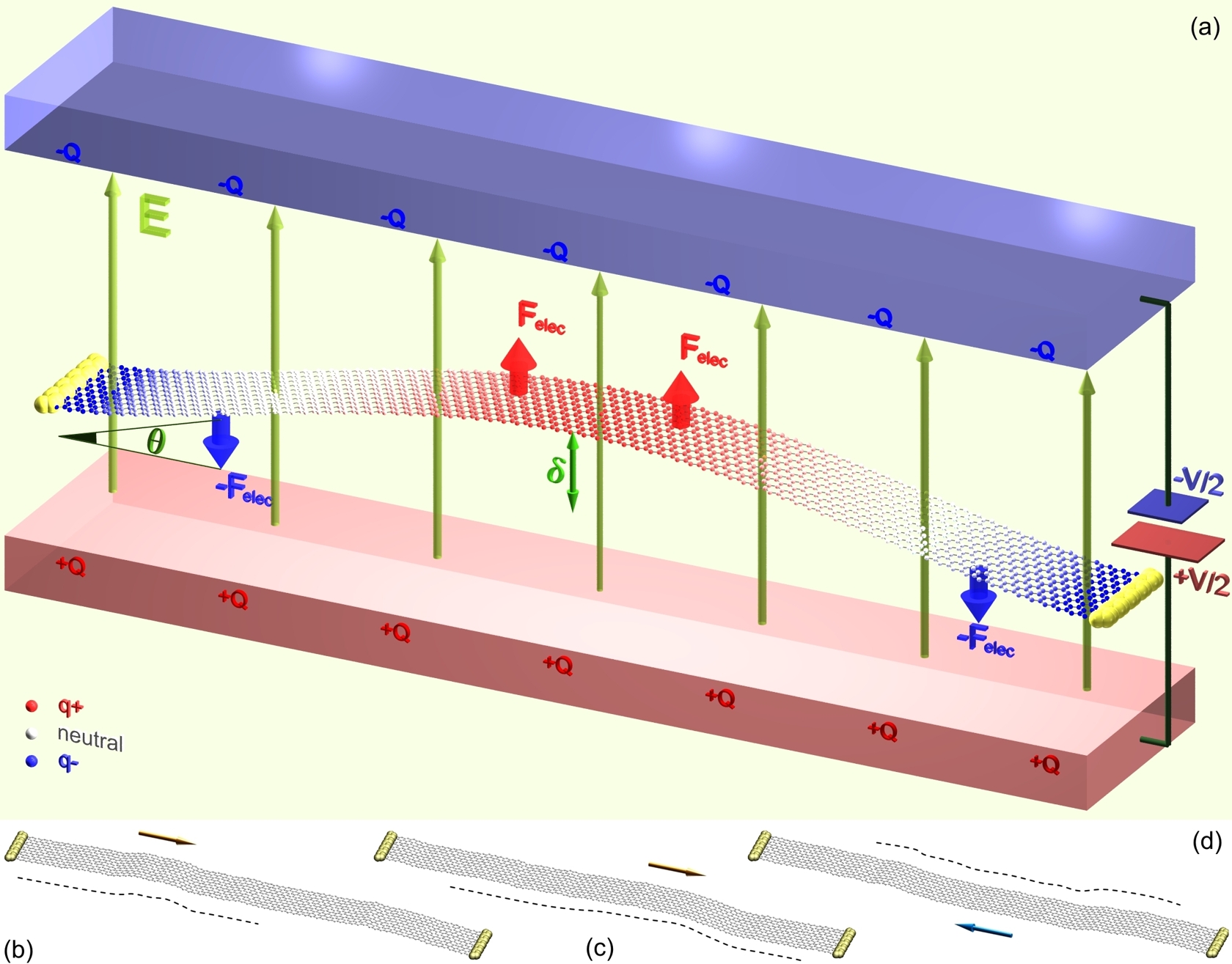}}
\caption{\label{fig:Schema}
(Color online) (a) Topographic diagram of charge distribution on a suspended GNR ($L=20$nm) in a transverse electric field $\bm{E}$ from molecular simulations. The color scale of atoms is proportional to the induced-charge density. $\bm{F}_{elec}$ and $-\bm{F}_{elec}$ stand for the electrostatic force arising from the interaction between the charges and the field, which make the GNR deflected. $\delta$ and $\theta$ stand for the maximal amplitude and angle of deflection, respectively. The electric field can be generated by applying a gate voltage between parallel capacitors. (b-d) Representative atomic configurations of the graphene at room temperature from MD simulations before the field is applied. The dashed lines represent the cross section shapes and the arrows show the direction of elastic wave propagation.} 
\end{figure}

Technically, no deformation will take place if an electric field is applied perfectly perpendicular to the graphene surface, because the induced dipole is already parallel to the field direction hence the induced moment of electrostatic force is zero \cite{Guo2003b}. However, this ideal situation could be seldom attained due to previously predicted intrinsic height fluctuations \cite{Fasolino2007} and edge stresses \cite{Shenoy2008} in graphene. During verification of these predictions using MD simulations based on the AIREBO potential, we have observed interesting elastic wave propagation on the surface of graphene at room temperature (see Fig. \ref{fig:Schema} (b-d)). The typical speed and maximal vertical amplitude of this wave propagation in the graphene are found to be about $2$ km/s and $0.2-0.4$nm, respectively. The elastic waves (so called intrinsic ripples) make the shape of GNRs more or less naturally curved. This natural curvature provides possibility for an electric field to shift negative and positive charges to opposite directions in graphene.  
 
To show the physical principle of electric deflection, we depict in Fig. \ref{fig:Schema} (a) the profile of polarized charges in suspended graphene by a transverse electric field (computational details can be found elsewhere \cite{ZW2009}). We see that positive and negative charges in a GNR are shifted to its top (center) and bottom (side), respectively. Two pairs of opposite forces $F_{elec}$ arise from the electrostatic interaction between the field and polarized charges, and form a bending moment $\bm{M}$ acting on each half graphene. Mirror symmetry of the system indicates that the GNR can equally be deflected downward, depending on the initial curvature of graphene at the moment when the gate voltage is applied. 

\begin{figure}[ht]
\centerline{\includegraphics[width=13cm]{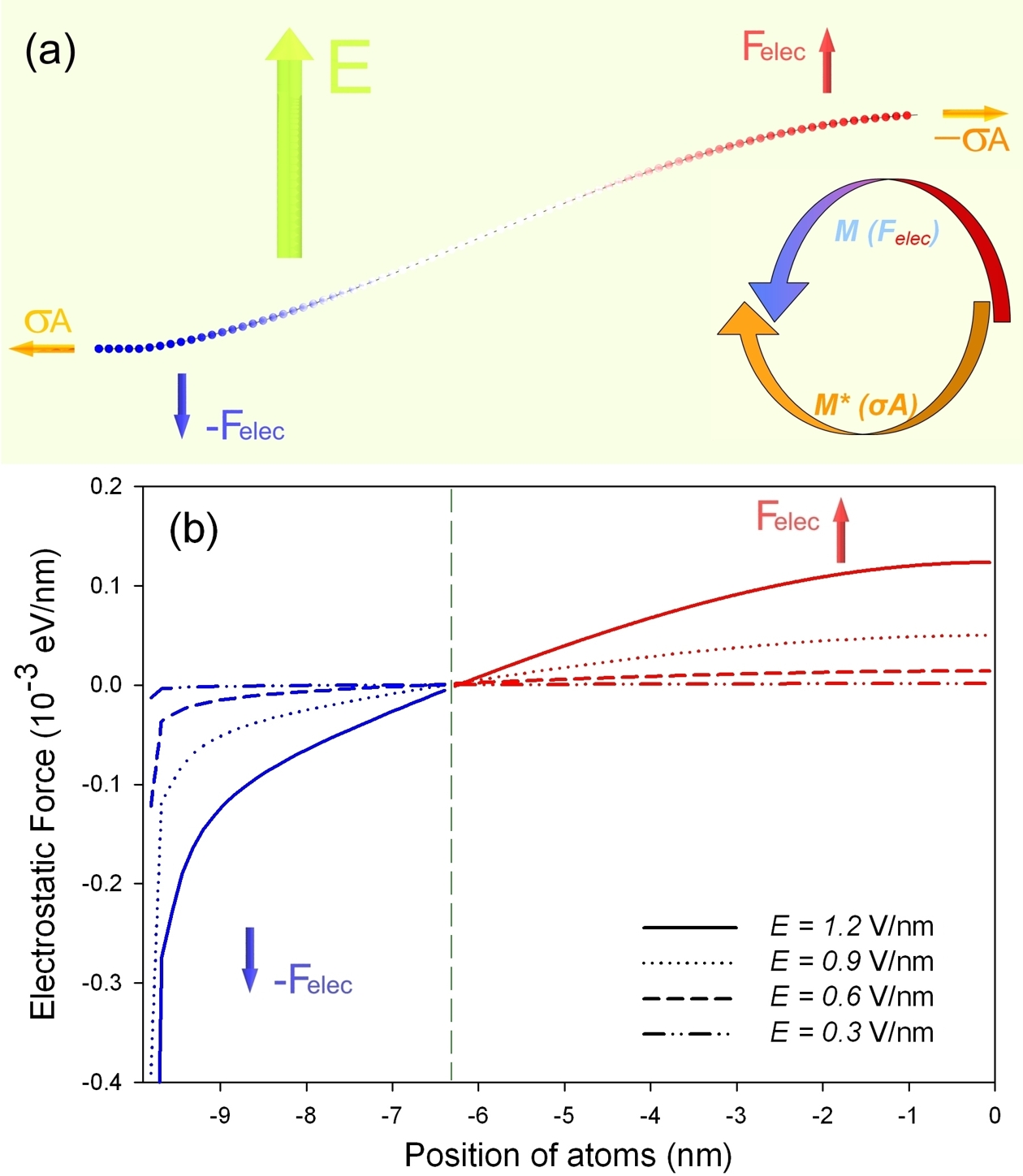}}
\caption{\label{fig:force}
(Color online) (a) Schematic of force/moment balance on a suspended GNR ($L/2 \approx 10$nm, from the left end to middle) in a transverse electric field $\bm{E}$. $\sigma A$ stands for a force pair of internal stress arising from the deflection of graphene. $\bm{M}$ and $\bm{M}^{*}$ correspond respectively to the moment of electric force and internal stress. (b) Profile of electric force acting on atoms in different positions along a half of the graphene with different field strengths $E$.
}
\end{figure}

The key to modeling this electromechanical behavior is to understand the force equilibrium in graphene. We consider the force balance in a half GNR as shown in Fig. \ref{fig:force} (a), which highlights the membrane-like character of graphene's mechanical properties. The formation of a moment of electric force can be understood by results plotted in Fig. \ref{fig:force} (b). This force profile is quite different from a commonly used assumption of a uniform electric force distribution in simplified calculations \cite{Fogler2008,Jonsson2005}. In general, the amplitude of driving electric force should be proportional to either the polarizability of GNR ($F_{elec} \propto \alpha$), or the square of field strength ($F_{elec} \propto E^{2}$) \cite{kozinsky-06}. Since the longitudinal polarizability of graphene is usually much larger than the transverse one  \cite{Brothers2005} ($\alpha_{//} >> \alpha_{\bot}$), $\bm{M}$ created in each half of the graphene by a electric force distribution (Fig. \ref{fig:force} (b)) can be calculated analytically as 

\begin{equation}
\label{Eq:F}
M = E^{2} \alpha_{//} \sin {\theta} \cos {\theta} 
\end{equation}  

\noindent where $\theta$ stands for the angle of deflection (Fig. \ref{fig:Schema} (a)). In our earlier work \cite{Wang200905}, it has been shown that $\alpha_{//}$ of a GNR is roughly proportional to its width $w$ or the square of its length $L^{2}$ when the graphene sheet is not too small. Thus, for the system that we study here, Eq. \ref{Eq:F} leads to
  
\begin{equation}
\label{Eq:F2}
M = E^{2} B w (\frac{L}{2})^{2}  \sin {\theta} \cos {\theta}
\end{equation}  

\noindent where $B=0.069$nm is a constant related to the dielectric constant of graphene. 

We now estimate the moment of membrane force $M^{*}$ induced by the graphene shape change during the deflection. Considering the fact that graphene's stiffness in its atomic plane is about 30 times higher than that in the perpendicular direction \cite{Bosak2007,Lee2008}, the internal stress in a curved GNR can be supposed to be mainly due to its axial deformation $\epsilon$, which can be approximated as : $\epsilon = \sigma /Y \approx 1/cos(\theta)-1$. Thus, $M^{*}$ can be expressed as

\begin{equation}
\label{Eq:F3}
M^{*} \approx -AY \frac{L}{2} \tan{\theta} (\frac{1}{ \cos{\theta}}-1)
\end{equation}  

\noindent where $Y$ is Young's modulus and $A$ is the cross sectional area. Here we use the value of their product $AY \approx 2120 w$ eV/nm recently measured from nanoindentation experiments \cite{Lee2008}. This value is in good agreement with that from theoretical calculations \cite{Shenoy2008} for GNRs with either armchair or zigzag edges. With the aid of moment balance in graphene $M=-M^{*}$ (shown in Fig. \ref{fig:force} (a)), and by combining Eqs. \ref{Eq:F2} and \ref{Eq:F3}, we obtain the governing equation of electrostatic deflection of suspended graphene as follows: 

\begin{equation}
\label{Eq:F4}
E = \sqrt{\frac{C}{L}(\frac{1}{\cos^{3}{\theta}}-\frac{1}{\cos^{2}{\theta}})}
\end{equation}  

\noindent where $C=2AY/Bw=60653 eV/nm^{2}$ is a constant. Since $w$ has been eliminated from this equation, we can first conclude that the deflection of a GNR is independent of its width. We note that increase of $\cos{\theta}$ with decreasing $w$ can be expected for narrow GNRs due to the effects of depolarization and edge states. However, this variation can be neglected for the size of GNRs usually reported in experiments. This equation also suggests that, for a given deformation angle $\theta$, required field strength $E$ decreases linearly with increasing the square root of graphene length $\sqrt{L}$. It is important to note that, since the parameters of our charge-dipole model were fitted for metallic $sp^{2}$ system, Eq. \ref{Eq:F4} may \textit{not be useful} for very narrow graphene sheets (with the minimum lateral dimension $<6$ nm) which have large band gaps due to edge states \cite{Ritter2009,Son2006b}.  

We now consider a particular case of small deflection (typically $\theta< \pi / 12$), for mainly covering experimentally reported resonance amplitude \cite{Bunch2007,Sazonova2004}. In such a case Eq. \ref{Eq:F4} implies that $\theta$ is roughly proportional to $E$ since geometric simplification gives $\cos^{-3}{\theta} - \cos^{-2}{\theta} = \tan{(0.5\theta)} \tan{\theta}\cos^{-2}{\theta}$. The maximal deflection $\delta$ (Fig. \ref{fig:Schema} (a)) can be approximately calculated as: $\delta = L\tan{\theta}/2$. Thus, the case of small deflection from Eq \ref{Eq:F4} leads to 

\begin{equation}
\label{Eq:F5}
\delta \approx E \sqrt{\frac{L^{3}}{2C}} .
\end{equation}  

\begin{figure}[ht]
\centerline{\includegraphics[width=13cm]{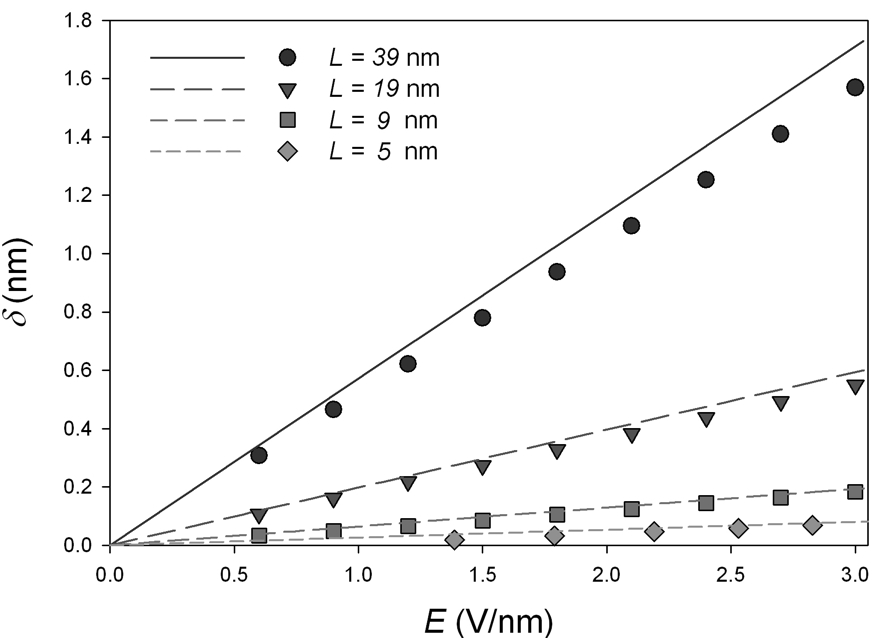}}
\caption{\label{fig:Filedstrength}
$\delta$ \textit{vs} $E$ for GNRs with different lengths $L$. The symbols represent simulation data and the lines stand for those predicted by analytical model using Eq.\ref{Eq:F4} (unit conversion: $1$V/nm $ \approx 0.833 \sqrt{eV/nm^{3}}$). }
\end{figure}

The analytical model presented in Eq. \ref{Eq:F5} is validated through series of molecular simulations for GNRs of different lengths. Results shown in Fig.\ref{fig:Filedstrength} remark a quantitative agreement between analytical prediction and simulation data. It can be seen that $\delta$ roughly follows a linear relationship with $E$ and rapidly increases with $L$, as predicted by Eq. \ref{Eq:F5}. The slight difference between the slopes of the curves is supposed to be mainly due to geometric approximations and the difference between experimentally measured Young's modulus and the one predicted from simulations.  

We now estimate the possibility of graphene's free oscillation induced by this electrostatic deflection by means of MD. In such simulations we apply a transverse electric field to deflect a suspended GNR, then the field is removed for inducing free oscillation of the GNR around its equilibrium position \cite{Sazonova2004}. We simulated the vibration of graphene during a period of time (some ns), and we counted the number of oscillation from the output of the simulations. We found that the GNR oscillates with different harmonics, depending on its length. Only short GNRs ($L < 12$ nm) were observed to oscillate at fundamental harmonic and other GNRs oscillate at high-order ones. This is because the field-induced resonance in a GNR is a combination of transverse and longitudinal waves (see MD simulation video-recording \cite{movie}), while the longitudinal wave has no space to propagate in very short graphene hence only the fundamental harmonic was observed. We note that the vibration mode and frequency can also vary with the initial tension from supports due to different nature of fixation \cite{Garcia-Sanchez2008}.   

\begin{figure}[ht]
\centerline{\includegraphics[width=13cm]{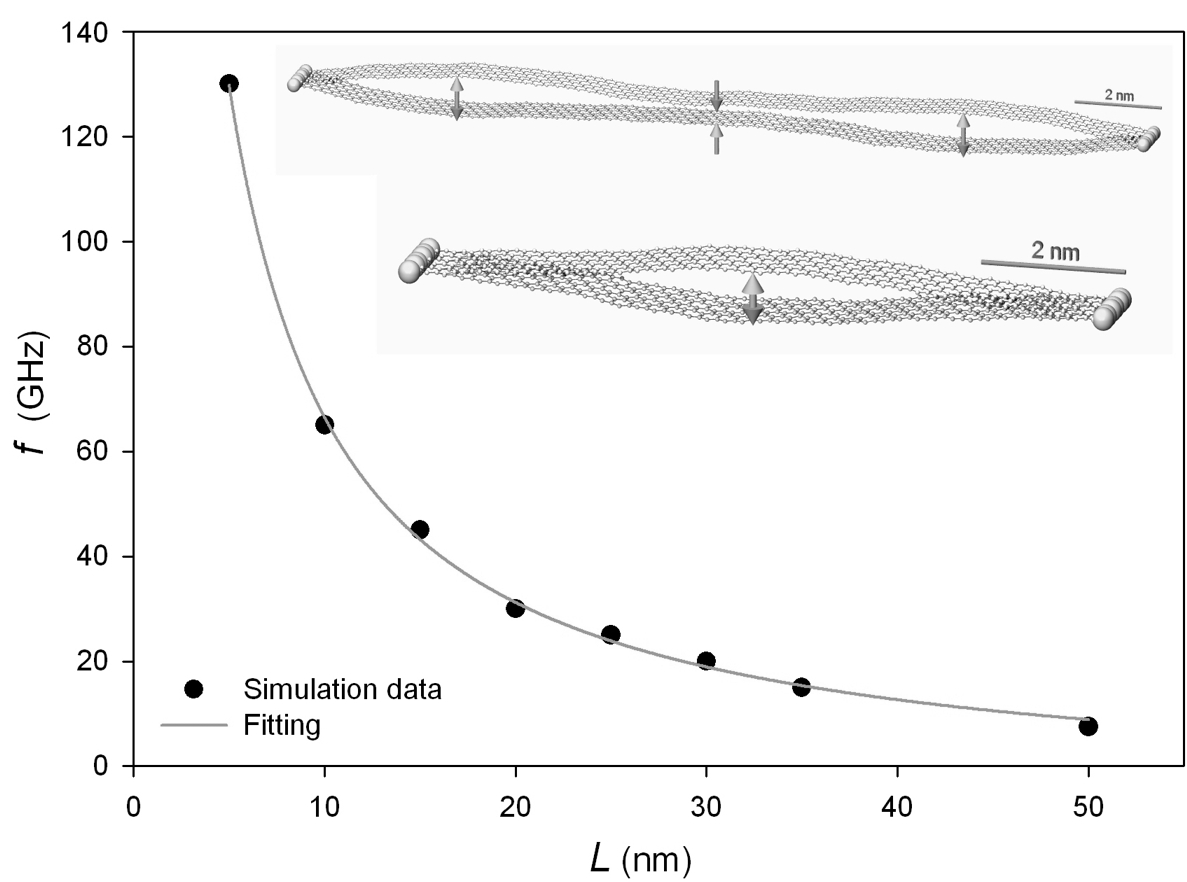}}
\caption{\label{fig:frequency}
Resonance frequency $f$ of GNRs of different lengths $L$. Inset shows that GNRs oscillate with different harmonics modes. The circles represent simulation data and the curve stands for a best-fitting equation: $f=-6.532+778.1L^{-1}-482.4L^{-2}$.
}
\end{figure}

We measured the natural frequency $f$ of GNRs with different sizes. From Fig. \ref{fig:frequency} we can see that $f$ decreases rapidly with increasing $L$. This length dependence is comparable to those found for single-walled CNTs \cite{Li2003b}. In general, graphene's free oscillation is tunable by applying an alternating voltage, by which the mechanical vibration can be either enhanced or attenuated in terms of adjusting the AC frequency \cite{Sazonova2004}. Theoretically speaking, the suspended structure could even be destroyed if the gate frequency is close to the natural frequency of the graphene. 

In conclusion, we have demonstrated that a suspended GNR can be deflected by applying a transverse electric field. The strong correlation between the field strength, graphene size and induced deflection has been concluded in an analytical model, which is validated via simulations. It was found that the deflection of GNRs is roughly proportional to the field strength for small deformations, and increases with the graphene length. The graphene's resonance frequency induced by this electrostatic deflection is found to decrease dramatically with the graphene length. These results suggest new potential applications of GNRs in electromechanical resonators or oscillators, which allow direct conversion from electric potential energy to mechanical energy in nanoscale. 

\section*{Acknowledgments}

We gratefully thank S. J. Stuart and R. Langlet for their help in the implementation of our computational code. D. Stewart, L. Henrard, A. Mayer, M. Devel and W. Ren are acknowledged for useful discussions.


\newpage

\section{Figures}

\newpage

\newpage

\newpage

\end{document}